\documentclass[12pt]{article}
\usepackage{epsfig}
\usepackage{axodraw}
\usepackage{color}
\usepackage{colordvi}

\newcommand{\CLra}{{C_{L,q}^{N(1)}\over C_{L,g}^{N(1)}}}

\newcommand{\CLrb}{{C_{L,q}^{N(2)} \over C_{L,g}^{N(1)}}}
\newcommand{\CLrc}{{C_{L,g}^{N(2)} \over C_{L,g}^{N(1)}}}
\newcommand{\Gqqa}{\gamma_{qq}^{N(0)}}
\newcommand{\Gqga}{\gamma_{qg}^{N(0)}}
\newcommand{\Ggqa}{\gamma_{gq}^{N(0)}}
\newcommand{\Ggga}{\gamma_{gg}^{N(0)}}
\newcommand{\Gqqb}{\gamma_{qq}^{N(1)}}
\newcommand{\Gqgb}{\gamma_{qg}^{N(1)}}
\newcommand{\Ggqb}{\gamma_{gq}^{N(1)}}
\newcommand{\Gggb}{\gamma_{gg}^{N(1)}}

\newcommand{\Cq}{C_{2,q}^{N(1)}}
\newcommand{\Cg}{C_{2,g}^{N(1)}}
\newcommand{\bb}{{\beta_1 \over \beta_0}}

\newcommand{\bea}{\begin{eqnarray}}
\newcommand{\bq}{\begin{equation}}
\newcommand{\eea}{\end{eqnarray}}
\newcommand{\eq}{\end{equation}}

\newcommand{\lsim}{\raisebox{-0.07cm}
{$\, \stackrel{<}{{\scriptstyle\sim}}\, $}}
\newcommand{\gsim}{\raisebox{-0.07cm}
{$\, \stackrel{>}{{\scriptstyle\sim}}\, $}}
\newcommand\pubnumber{DESY 01-004}
\newcommand\pubdate{\today}
\newcommand\hepnumber{hep-ph/0101235}

\def\csumb{$^a$~DESY Zeuthen, Platanenallee 6, \\
D-15738 Zeuthen, Germany\\
     $^b$~Harish--Chandra Research Institute,
     Chhatnag Road, Jhusi, \\
Allahabad, 211019, India \\
$^c$Instituut-Lorentz,
Universiteit Leiden,
P.O. Box 9506, \\
2300 HA Leiden, The Netherlands.
}
\def\support{\footnote{Work supported in part by EU contract 
FMRX-CT98-0194 (DG 12-MIHT).}}
\def\pres{\footnote{Presented by J. Bl\"umlein}}

\def\Title#1{\begin{center} {\Large\bf #1 } \end{center}}
\def\Author#1{\begin{center}{ \sc #1} \end{center}}
\def\Address#1{\begin{center}{ \it #1} \end{center}}

\newcommand\pubblock{\rightline{\begin{tabular}{l} \pubnumber\\
         \pubdate\\ \hepnumber \end{tabular}}}
\newenvironment{Abstract}{\begin{quotation}  }{\end{quotation}}
\newenvironment{Presented}{\begin{quotation} \begin{center} 
             Presented at the\end{center}
      \begin{center}\begin{large}}{\end{large}\end{center} \end{quotation}}
\def\Acknowledgments{\bigskip  \bigskip \begin{center}
          \large\bf Acknowledgments\end{center}}

\makeatletter
\def\section{\@startsection{section}{0}{\z@}{5.5ex plus .5ex minus
 1.5ex}{2.3ex plus .2ex}{\large\bf}}
\def\subsection{\@startsection{subsection}{1}{\z@}{3.5ex plus .5ex minus
 1.5ex}{1.3ex plus .2ex}{\normalsize\bf}}
\def\subsubsection{\@startsection{subsubsection}{2}{\z@}{-3.5ex plus
-1ex minus  -.2ex}{2.3ex plus .2ex}{\normalsize\sl}}

\renewcommand{\@makecaption}[2]{%
   \vskip 10pt
   \setbox\@tempboxa\hbox{\small #1: #2}
   \ifdim \wd\@tempboxa >\hsize     % IF longer than one line:
       \small #1: #2\par          %   THEN set as ordinary paragraph.
     \else                        %   ELSE  center.
       \hbox to\hsize{\hfil\box\@tempboxa\hfil}
   \fi}

 \def\citenum#1{{\def\@cite##1##2{##1}\cite{#1}}}
\newcount\@tempcntc
\def\@citex[#1]#2{\if@filesw\immediate\write\@auxout{\string\citation{#2}}\fi
  \@tempcnta\z@\@tempcntb\m@ne\def\@citea{}\@cite{\@for\@citeb:=#2\do
    {\@ifundefined
       {b@\@citeb}{\@citeo\@tempcntb\m@ne\@citea\def\@citea{,}{\bf ?}\@warning
       {Citation `\@citeb' on page \thepage \space undefined}}%
    {\setbox\z@\hbox{\global\@tempcntc0\csname b@\@citeb\endcsname\relax}%
     \ifnum\@tempcntc=\z@ \@citeo\@tempcntb\m@ne
       \@citea\def\@citea{,}\hbox{\csname b@\@citeb\endcsname}%
     \else
      \advance\@tempcntb\@ne
      \ifnum\@tempcntb=\@tempcntc
      \else\advance\@tempcntb\m@ne\@citeo
      \@tempcnta\@tempcntc\@tempcntb\@tempcntc\fi\fi}}\@citeo}{#1}}
\def\@citeo{\ifnum\@tempcnta>\@tempcntb\else\@citea\def\@citea{,}%
  \ifnum\@tempcnta=\@tempcntb\the\@tempcnta\else
  {\advance\@tempcnta\@ne\ifnum\@tempcnta=\@tempcntb \else\def\@citea{--}\fi
    \advance\@tempcnta\m@ne\the\@tempcnta\@citea\the\@tempcntb}\fi\fi}
\makeatother

\def\beq{\begin{equation}}
\def\eeq#1{\label{#1}\end{equation}}
\def\eeqn{\end{equation}}

\newenvironment{Eqnarray}%
   {\arraycolsep 0.14em\begin{eqnarray}}{\end{eqnarray}}
\def\beqa{\begin{Eqnarray}}
\def\eeqa#1{\label{#1}\end{Eqnarray}}
\def\eeqan{\end{Eqnarray}}

\let\bar=\overbar

\def\Dslash{\not{\hbox{\kern-4pt $D$}}}
\def\dslash{\not{\hbox{\kern-2pt $\del$}}}

\def\msb{{\bar{\ssstyle M \kern -1pt S}}}

\def\lsim{\mathrel{\raise.3ex\hbox{$<$\kern-.75em\lower1ex\hbox{$\sim$}}}}
\def\gsim{\mathrel{\raise.3ex\hbox{$>$\kern-.75em\lower1ex\hbox{$\sim$}}}}

\begin{document}
\begin{titlepage}
\sloppy
\pubblock

\vfill
\def\thefootnote{\fnsymbol{footnote}}
\Title{The Drell-Yan-Levy Relation: \\ \vspace{1mm}
{\boldmath $ep$} vs {\boldmath
$e^+e^-$} Scattering to {\boldmath $O(\alpha_s^2)\pres\support$}}
\vfill
\Author{$^a$~J. Bl\"umlein, $^b$~V. Ravindran, and $^c$~W.L. van Neerven}
\Address{\csumb}
\vfill
%%%%%%%%%%%%%%%%%%%%%%%%%%%%%%%%%%%%%%%%%%%%%%%%%%%%%%%%%%%%%%%%%%%%%%%%%%%
\begin{Abstract}
\noindent
We study the validity of a relation by Drell, Levy and Yan (DLY) 
connecting the deep inelastic structure (DIS) functions and the 
single-particle fragmentation functions in $e^+e^-$ annihilation 
which are defined in the spacelike ($q^2<0$) and timelike ($q^2>0$) 
regions, with respect to physical evolution kernels for the two 
processes to $O(\alpha_s^2)$. We also comment on a relation proposed 
by Gribov and Lipatov.
\end{Abstract}
%%%%%%%%%%%%%%%%%%%%%%%%%%%%%%%%%%%%%%%%%%%%%%%%%%%%%%%%%%%%%%%%%%%%%%%%%%%
\vfill
\begin{Presented}
5th International Symposium on Radiative Corrections \\ 
(RADCOR--2000) \\[4pt]
Carmel CA, USA, 11--15 September, 2000
\end{Presented}
\vfill
\end{titlepage}
\def\hefootnote{\arabic{footnote}}
\setcounter{footnote}{0}
\newpage
%===============================================================================
\section{The DLY-Relation}
%===============================================================================

\vspace{1mm}
\noindent
Right after the discovery of the partonic structure of nucleons the
question arose,
whether at large virtualities of the exchanged photon
the hard processes {$e^- p \rightarrow e^- X$}
and
{$e^+ e^- \rightarrow \overline{p} X$} are related by crossing from the
$t-$ to the $s-$channel,~\cite{BAS1,BAS2}.
%-------------------------------------------------------------------------
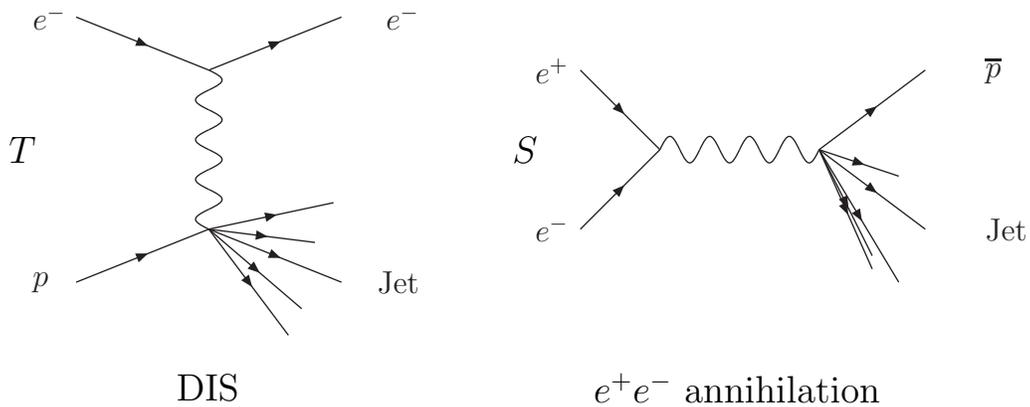
\begin{center} 
%--------------------         ep -> eX
\setlength{\unitlength}{0.2mm}
\begin{picture}(800,230)(0,0)
\SetColor{Black}
\ArrowLine(50,100)(100,80)
\SetColor{Black}
\ArrowLine(100,80)(150,100)
\SetColor{Black}
\Photon(100,80)(100,20){5}{4}
\SetColor{Black}
\ArrowLine(50,0)(100,20)
\SetColor{Black}
\ArrowLine(100,20)(150,0)
\SetColor{Black}
\ArrowLine(100,20)(130,-20)
\SetColor{Black}
\ArrowLine(100,20)(147, 30)
\SetColor{Black}
\ArrowLine(100,20)(140, 15)
\SetColor{Black}
\ArrowLine(100,20)(135,-10)
%-------------------------------------------------------------------------
%-------------------------------------------------------------------------
%--------------------         e+e- -> anti p X
\SetColor{Black}
\ArrowLine(240,80)(270,50)
\SetColor{Black}
\ArrowLine(240,20)(270,50)
\SetColor{Black}
\Photon(270,50)(330,50){5}{4}
\SetColor{Black}
\ArrowLine(330,50)(370,80)
\SetColor{Black}
\ArrowLine(330,50)(370,20)
\SetColor{Black}
\ArrowLine(330,50)(360,0)
\SetColor{Black}
\ArrowLine(330,50)(360,40)
\SetColor{Black}
\ArrowLine(330,50)(350,10)
\SetColor{Black}
\ArrowLine(330,50)(350,5)
\setlength{\unitlength}{1pt}
\SetColor{Black}
\Text(400,80)[r]{       $\overline{p}$}
\SetColor{Black}
\Text(220,80)[l]{       $e^+$}
\SetColor{Black}
\Text(220,20)[l]{       $e^-$}
\SetColor{Black}
\Text(410,20)[r]{       Jet}
%-------------------------------------------------------------------------
\SetColor{Black}
\Text(30,0)[l]{       $p$}
\SetColor{Black}
\Text(30,100)[l]{       $e^-$}
\SetColor{Black}
\Text(180,100)[r]{       $e^-$}
\SetColor{Black}
\Text(180,0)[r]{       Jet}
%-------------------------------------------------------------------------
\Text(100,-40)[]{\textcolor{black}{\large DIS}}
\Text(300,-40)[]{\textcolor{black}{\large $e^+e^-$ annihilation}}
%-------------------------------------------------------------------------
\Text(30,50)[]{\textcolor{black}{\large $T$}}
\Text(220,50)[]{\textcolor{black}{\large $S$}}
%-------------------------------------------------------------------------
\end{picture} 
\end{center}
%-------------------------------------------------------------------------
%-------------------------------------------------------------------------

\vspace{18mm}
\noindent
For the two--fold differential scattering cross sections for both 
processes,
%-------------------------------------------------------------------------
\begin{eqnarray}
\frac{d^2 \sigma}{dx dQ^2} \sim L_{\mu\nu} W^{\mu\nu}~,
\end{eqnarray}
%-------------------------------------------------------------------------
one may express channel crossing by the hadronic tensors $W_{\mu\nu}$
as done by Drell, Levy, and Yan~\cite{BAS1}
%-------------------------------------------------------------------------
\\
\begin{equation}
W_{\mu\nu}^{(S)}(q,p) = - W_{\mu\nu}^{(T)}(q,-p)~.
\end{equation}
%-------------------------------------------------------------------------
At that time partons were assumed as fermionic particles, interacting
via (pseudo) scalars, with $\delta(1-z)$-sources, where $z$ denotes the 
longitudinal momentum fraction. Considering only ladder graphs at lowest
order, such a crossing could be envisaged for the whole hadronic tensor.

Viewing these reactions in QCD, the picture changes. The sources of
the partons are extended non-perturbative distributions with 
$z~\epsilon~[0,1]$, about which perturbative QCD cannot make a statement,
even resumming whole classes of graphs. However, at large scales of $Q^2$
both processes factorize into the parton densities and perturbative
evolution kernels, which rule the $Q^2$ behaviour.  The question of
crossing from the $t-$ to the $s-$ channel can thus be modified in
studying it for the {\it factorized} evolution kernels at the one
side within {\it perturbative QCD} and leaving the related question for
the non-perturbative sources to {\it Lattice Gauge Theory}.

The scaling variables describing deep inelastic scattering at the one
side and hadron fragmentation on the other side are $x_B$ and $x_E$,
%------------------------------------------------------------------------
\begin{eqnarray}
x_B &=& \frac{Q^2}{2 p.q}
,~~0 \leq x_B \leq 1~~~~   {{\rm DIS}}         \\ %
x_E &=& \frac{2 p.q}{Q^2}
,~~0 \leq x_E \leq 1~~~~   {e^+e^-}~   {{\rm annihilation}}~.
\end{eqnarray}
%------------------------------------------------------------------------
The point $x =1$ connects both domains and is usually a singular point.
One may now calculate QCD evolution kernels for both domains. The central
question of the present paper is, what  are the  conditions to
{\it continue} the kernel obtained in one domain into that of the other.
In general one cannot expect to find an analytic continuation in an
arbitrarily chosen factorization scheme in which the process independent
splitting functions are evaluated. However, one may form {\it
physical evolution kernels}, in both domains, which are scheme--invariant
and study their crossing behaviour. Thus the above question is directed
to the connection of the physical evolution behaviour of {\it observables}
as the structure and fragmentation functions. In the present paper we 
study this relation up to $O(\alpha_s^2)$.
Early related investigations (partly before the advent of QCD) were
performed in Refs.~\cite{DL}--\cite{BUKH}, and more recently in
Refs.~\cite{cfp}--\cite{BRV1}.

%===============================================================================
\section{Scheme--invariant Evolution Equations}
%===============================================================================

\vspace{1mm}
\noindent
To investigate the crossing behaviour for the evolution kernels of
structure and fragmentation functions we first derive physical
evolution equations. The twist-2 contributions to these functions
can be expressed in the form
%------------------------------------------------------------------------
\begin{eqnarray}
\label{eqFA}
F_i(x,Q^2)=\sum_{l=q,g} \left ( C_{i,l} \Big(\alpha_s(\mu_r^2),
{Q^2 \over \mu_f^2}, {\mu_f^2 \over \mu_r^2}\Big ) \otimes 
f_l\Big(\alpha_s(\mu_r^2),{\mu_f^2 \over \mu^2},{\mu_f^2 
\over \mu_r^2}\Big) \right )(x),
\end{eqnarray}
%------------------------------------------------------------------------
where $C_{i,l}$ denote the Wilson coefficients, $f_l$ the parton
densities and $\otimes$ is the Mellin convolution. $\mu_f^2$ and 
$\mu_r^2$ are the factorization and renormalization scales, respectively.
Beyond leading order the parton 
densities and Wilson coefficients obey factorization scheme--dependent 
evolution equations and are thus no observables. Their dependence
on $\mu_f^2$ can, however, be eliminated in expressing the non--singlet
and singlet parton densities via physical observables, the scale
dependence of which is finally studied. For the singlet case one obtains
%------------------------------------------------------------------------
\begin{eqnarray}
\label{eqevo2}
{\partial \over \partial t}\left( \begin{array}{c}
F^N_A\\
{F^N_B}
\end{array} \right)= -{1 \over 4}
\left( \begin{array} {cc}
K^N_{AA} & K^N_{AB}\\
K^N_{BA} & K^N_{BB}
\end{array} \right)
\left( \begin{array}{c}
F^N_A\\
{F^N_B}
\end{array} \right)~.
\end{eqnarray}
%------------------------------------------------------------------------
Here the evolution kernels $K^N_{IJ}$ written in Mellin--moment space
are no longer process--independent quantities for the evolution of the
pair of observables $\{F_A^N,F_B^N\}$, but scheme--independent
quantities. Eq.~(\ref{eqevo2}) refers to the evolution variable  
$t= - (2/\beta_0) \times \newline
\ln(a_s(Q^2)/a_s(Q^2_0))$.
$\beta_0$ is the lowest order $\beta-$function, and
$a_s(Q^2) = \alpha_s(Q^2)/(4\pi)$. The physical evolution kernels read
%------------------------------------------------------------------------
\begin{eqnarray}
\label{KERN1}
K^N_{IJ}=\Bigg [ -4 \frac{\partial C^N_{I,m}(t) }
{\partial t } \left (C^N\right )_{m,J}^{-1}(t)
 - {\beta_0 a_s(Q^2) \over  \beta(a_s(Q^2))}
C^N_{I,m}(t) \gamma^N_{mn}(t)
\left (C^N \right )_{n,J}^{-1}(t)\Bigg ]~,
\end{eqnarray}
%------------------------------------------------------------------------
with $\beta(a_s)$ the $\beta-$function and
%------------------------------------------------------------------------
\begin{eqnarray}
\label{KERN2}
K^N_{IJ} = \sum_{n=0}^{\infty} 
a_s^n(Q^2) \left (K^N \right )_{IJ}^{(n)}~.
\end{eqnarray}
%------------------------------------------------------------------------
One easily sees that the kernels Eq.~(\ref{KERN1}) are very difficult
to obtain in $x$--space, due to the inverse coefficient functions
to be evaluated. Instead they take a simple form for the 
Mellin-transforms. The transformed coefficient functions are needed in 
{\it analytic form} in $N$, which are usually polynomials out of multiple
alternating and non--alternating harmonic
sums~\cite{HARS1,HARS2}. These expressions have to be analytically
continued to complex values of $N$. It turns out that all Wilson
coefficients to $O(\alpha_s^2)$ can be expressed by at most 26 basic
functions of complex $N$, the analytic continuations of  which can
be found in Ref.~\cite{HARS3}.

Possible choices for the observables $\{F_A^N,F_B^N\}$ are
${F_2,\partial F_2/\partial t}$, ${g_1,\partial g_1/\partial t}$, and
${F_2,F_L}$. Here we denote by $F_i$ and $g_i$ the respective
unpolarized and polarized structure {\it and} fragmentation functions.
The physical evolution kernels, as obtained from the anomalous dimensions
and coefficient functions, read~:

\vspace{1mm} \noindent
\begin{center}
{\bf System:~{\boldmath $F_2, \partial F_2/\partial t$}}\\
\end{center}
{Leading Order}~\cite{FP}:\\
%------------------------------------------------------------------------
\begin{eqnarray}
K_{22}^{N(0)}&=&0   \nonumber\\
K_{2d}^{N(0)}&=&-4 \nonumber\\
K_{d2}^{N(0)}&=&             {{1 \over 4} \Bigg(\Gqqa \Ggga-\Gqga
\Ggqa\Bigg)}   \nonumber\\
K_{dd}^{N(0)}&=&             {\Gqqa+\Ggga }
\end{eqnarray}
%------------------------------------------------------------------------
{Next-to-Leading Order}~\cite{FP}:\\
%------------------------------------------------------------------------
\begin{eqnarray}
\label{eqG22}
K_{22}^{N(1)}&=& K_{2d}^{N(1)} = 0 \nonumber\\
K_{d2}^{N(1)}&=&{1 \over 4} \Bigg[\Ggga \Gqqb+\Gggb \Gqqa -\Gqgb 
\Ggqa -\Gqga \Ggqb\Bigg]
\nonumber\\[2ex]
&& -{\beta_1 \over 2 \beta_0} \Bigg(\Gqqa \Ggga-\Ggqa \Gqga\Bigg)
\nonumber \\ & &
   +{\beta_0 \over 2} \Cq \Bigg(\Gqqa+\Ggga-2 \beta_0\Bigg)
\nonumber\\[2ex]
&&-{\beta_0 \over 2} {\Cg \over \Gqga} \Bigg[(\Gqqa)^2-\Gqqa\Ggga+2 
\Gqga\Ggqa-2 \beta_0 \Gqqa\Bigg]
\nonumber\\[2ex]
&&-{\beta_0 \over 2} \Bigg(\Gqqb-{\Gqqa \Gqgb \over \Gqga} \Bigg)
\nonumber\\
\end{eqnarray}
%===============================================================================
The same structures apply to ${g_1,\partial g_1/\partial t}$.

\vspace{1mm} \noindent
\begin{center}
{\bf System:~{\boldmath $F_2, \hat{F}_L$}}\\
\end{center}
For convenience we define
{$\hat{F}_L^{N} \equiv F_L^N/(a_s(Q^2) C_{L,g}^{N(1)})$}.\\
%------------------------------------------------------------------------
{Leading Order}~\cite{CAT}:\\
%------------------------------------------------------------------------
\begin{eqnarray}
\label{eqG11}
K_{22}^{N(0)}\!\!&=&\!\!\displaystyle{\Gqqa- \CLra \Gqga }
\nonumber\\
K_{2L}^{N(0)}\!\!&=&\!\!\displaystyle{\Gqga} 
\nonumber\\
K_{L2}^{N(0)}\!\!&=&\!\!\displaystyle{\Ggqa-\left( \CLra \right)^2 
\Gqga} \nonumber\\
K_{LL}^{N(0)}\!\!&=&\!\!\displaystyle{\Ggga+\CLra \Gqga} 
  +               {\CLra \Big(\Gqqa-\Ggga\Big)}
\end{eqnarray}
%------------------------------------------------------------------------
{Next-to-Leading Order}~{\cite{BRV1}: \\
%------------------------------------------------------------------------
\begin{eqnarray}
\label{eqG12}
K_{22}^{N(1)}&=&\Gqqb-\bb \Gqqa-\CLra \left(\Gqgb-\bb \Gqga\right) 
\nonumber\\ & &
+\CLra \Cg \Gqqa 
   -\CLra \Cg \Ggga 
\nonumber\\
&&-\left[\CLrb+\left(\CLra\right)^2 \Cg-\CLra \CLrc\right] \Gqga
\nonumber\\  & &
+\Cg \Ggqa 
+ 2 \beta_0 \left(\Cq-\CLra \Cg\right)  
\nonumber\\
%------------------------------------------------------------------------
K_{2L}^{N(1)}&=&\Gqgb-\bb \Gqga-\Cg (\Gqqa-\Ggga)+2 \beta_0 \Cg
\nonumber\\
&&+\left(\Cq+\CLra \Cg-\CLrc\right) \Gqga
\nonumber\\
%------------------------------------------------------------------------
K_{L2}^{N(1)}&=&\Ggqb-\bb \Ggqa+\CLra\left(\Gqqb-\bb\Gqqa \right)
\nonumber\\[2ex] 
&&-\left( \CLra \right)^2 \left(\Gqgb-\bb\Gqga\right)\nonumber
-\CLra 
\left(\Gggb-\bb\Ggga\right)
\nonumber\\[2ex] 
&&+\left[\CLrb -\CLra \Cq+\left(\CLra\right)^2 \Cg \right] 
\Gqqa
\nonumber\\[2ex] 
&&-\left[\left(\CLra \right)^3 \Cg+2 \CLra \CLrb - 
\left(\CLra\right)^2 \CLrc \right .
\nonumber\\[2ex]
&&\left. -\left( \CLra\right)^2 \Cq \right] \Gqga
  +2 \beta_0 \left( \CLrb- \CLra \CLrc \right)
\nonumber\\ & &
+\left( \CLra \Cg -\Cq +\CLrc \right) \Ggqa
\nonumber \\
&&-\left[ \CLrb +\left(\CLra\right)^2 \Cg -\CLra \Cq \right] \Ggga
\nonumber\\
%------------------------------------------------------------------------
K_{LL}^{N(1)}&=& \Gggb-\bb \Ggga+\CLra \left(\Gqgb-\bb \Gqga\right)
\nonumber\\[2ex]
&&-\CLra \Cg \Gqqa +\left[\CLrb-\CLra \CLrc \right. \nonumber\\ & &
\left.
+\left(\CLra\right)^2 
\Cg\right]\Gqga
\nonumber\\[2ex]
&&-\Cg \Ggqa + \CLra \Cg  \Ggga +2 \beta_0 \CLrc
\end{eqnarray}
%------------------------------------------------------------------------
\section{DLY--Relations for Evolution Kernels}
%------------------------------------------------------------------------

\vspace{1mm}
\noindent
The original crossing relation~\cite{BAS1}
%------------------------------------------------------------------------
\begin{eqnarray}
W_{\mu \nu}^T(q,p)=-W_{\mu \nu}^S(q,-p)
\end{eqnarray}
%------------------------------------------------------------------------
is modified to~\cite{BUKH}
%------------------------------------------------------------------------
\begin{eqnarray}
\label{dlyf}
F_i^{(S)}(x_B) = - (-1)^{2(s_1+s_2)} 
x_E F^{(T)}_i\left(\frac{1}{x_E}\right) \quad , \quad i=1,2,L\,
\end{eqnarray}
%------------------------------------------------------------------------
if particles of different spin $s_i$ contribute, again considering ladder
approaches to the problem with idealized sources. 

In the following we give the analytic continuation relations, cf.
Ref.~\cite{BRV1}, which yield the correct transformations for physical
kernels up to $O(\alpha_s^2)$. They read~:
%------------------------------------------------------------------------
\begin{eqnarray}
\label{tra1}
P(z) &\rightarrow& zP(1/z) \\
\label{tra2}
P_{ii} & \rightarrow& - P_{ii} \\
\label{tra3}
P_{qg}, P_{gq} & \rightarrow& {\rm cross~color~pre-factor} \\
\label{tra4}
\ln\left({Q^2}/{\mu_f^2}\right)_{\rm space-like} &\rightarrow&
\ln\left({Q^2}/{\mu_f^2}\right)_{\rm time-like} - i \pi~.
\\
\label{tra5}
\delta(1-z) &\rightarrow& - \delta(1-z) \\ 
\label{tra6}
\ln(1-z)    &\rightarrow& \ln(1-z) - \ln(z) + i\pi \\
\label{tra7}
\ln(\varepsilon) &\rightarrow& \ln(\varepsilon) + i\pi 
\end{eqnarray}
%------------------------------------------------------------------------
Due to Eq.~(\ref{tra5}),~Eq.~(\ref{dlyf}) does not hold for 
$x_B = x_E = 1,$ even in leading order, where the physical evolution
kernels are the splitting functions.

In next--to--leading order the differences between the analytically
continued space--like splitting functions and the time--like splitting
functions are~:
%------------------------------------------------------------------------
\begin{eqnarray}
\label{dlysplitqq}
\bar P_{qq}^{(1)S}-P_{qq}^{(1)T}&=&
-2 \beta_0 Z^{T(1)}_{qq}+Z^{T(1)}_{qg}
\otimes \bar P_{gq}^{(0)}-Z^{T(1)}_{gq}
\otimes \bar P_{qg}^{(0)}\,,
\nonumber\\
\label{dlysplitqg}
\bar P_{qg}^{(1)S}-P_{gq}^{(1)T}&=&
-2 \beta_0 Z^{T(1)}_{qg}+Z^{T(1)}_{qg}\otimes(\bar P_{gg}^{(0)}-
\bar P_{qq}^{(0)})
\nonumber \\ & &~~~~~~~~~~~~~~~~~~~
+\bar P_{qg}^{(0)} 
\otimes(Z^{T(1)}_{qq}-Z^{T(1)}_{gg})\,,
\nonumber\\
\label{dlysplitgq}
 \bar P_{gq}^{(1)S}-P_{qg}^{(1)T}&=&
-2 \beta_0 Z^{T(1)}_{gq}+Z^{T(1)}_{gq}\otimes(\bar P_{qq}^{(0)}-
\bar P_{gg}^{(0)})
\nonumber \\ & &~~~~~~~~~~~~~~~~~~~
+\bar P_{gq}^{(0)}
\otimes(Z^{T(1)}_{gg}-Z^{T(1)}_{qq})\,,
\nonumber\\
\label{dlysplitgg}
\bar P_{gg}^{(1)S}-P_{gg}^{(1)T}&=&
-2 \beta_0 Z^{T(1)}_{gg}+Z^{T(1)}_{gq}\otimes \bar P_{qg}^{(0)}
-Z^{T(1)}_{qg}\otimes \bar P_{gq}^{(0)}~,   
\end{eqnarray}
%------------------------------------------------------------------------
and do {\it not} vanish. Here we defined
%------------------------------------------------------------------------
\begin{eqnarray}
\label{eqZT}
Z^{T(1)}_{ij}&=&P_{ji}^{(0)} \cdot \Big(\ln(z) + a_{ji}\Big)~,
\nonumber
\end{eqnarray}
%------------------------------------------------------------------------
where for unpolarized scattering
%------------------------------------------------------------------------
\begin{eqnarray}
\label{eqaij}
a_{qq}=a_{gg}=0\quad ,\quad a_{qg}=-\frac{1}{2}\quad , \quad 
a_{gq}=\frac{1}{2}~,  
\end{eqnarray} 
%------------------------------------------------------------------------
and    for   polarized scattering
%------------------------------------------------------------------------
\begin{eqnarray}
\label{peqaij}
a_{ij}=0~,
\end{eqnarray}
%------------------------------------------------------------------------
cf. also~\cite{SV}.

The transformation of the NLO coefficient functions $C_{1,q(g)}$ and 
$C_{L,q(g)}$ are~:
%------------------------------------------------------------------------
\begin{eqnarray}
\label{dlyCTq1}
C_{1,q}^{(T)(1)}(z)+\left \{z\, C_{1,q}^{(S)(1)}\left({1 \over z}\right)
\right \}&=&Z^{(T)(1)}_{qq}
 \nonumber \\
\label{dlyCTg1}
{1 \over 2}\left [C_{1,g}^{(T)(1)}(z)-{C_F \over 2 N_f T_f}
\left \{2 z C_{1,g}^{(S)(1)}\left({1 \over z}\right)\right \}\right]&
=&Z^{(T)(1)}_{qg}~,
\end{eqnarray}
%------------------------------------------------------------------------
and
%------------------------------------------------------------------------
\begin{eqnarray}
\label{dlyCLq1}
C_{L,q}^{(T)(1)}(z)-{z \over 2}  C_{L,q}^{(S)(1)}\left({1 \over z}\right) &=& 0
\,,
\nonumber\\
\label{dlyCLg1}
{1 \over 2} \left [C_{L,g}^{(T)(1)}(z)+{C_F \over 2 N_f T_f} \left \{z \,
 C_{L,g}^{(S)(1)}\left({1 \over z}\right)\right \}\right] &=& 0~.
\end{eqnarray}
%------------------------------------------------------------------------
Finally the NNLO unpolarized longitudinal coefficient 
functions~\cite{ZNR} transform as~\cite{BRV1}~:
%------------------------------------------------------------------------
\begin{eqnarray}
\label{dlyCLq2}
\lefteqn{
\!\!\!\!C_{L,q}^{(T)(2)}(z)\!+\!\left \{-{z \over 2}\, C_{L,q}^{(S)(2)}
\left({1 \over z}\right) \right \}
       =}
  \nonumber\\ & &
\!Z^{(T)(1)}_{qq}\otimes {z \over 2} C_{Lq}^{(1)S}\left({1 \over z}\right)
%\nonumber\\[2ex]
   +Z^{(T)(1)}_{gq}\otimes {C_F \over 2 N_f T_f} \left \{-{z \over 2}\,
C_{L,g}^{(S)(1)}\left({1 \over z}\right) \right \} \,,
 \nonumber\\
%%%%R
\label{dlyCLg2}
\lefteqn{
{1 \over 2} \left[C_{L,g}^{(T)(2)}(z)\!+\!{C_F \over 2 N_f T_f} \left \{z 
 C_{L,g}^{(S)(2)}\left({1 \over z}\right)\right\}\right]
       =}
  \nonumber\\ & &
\!Z^{(T)(1)}_{qg}\otimes {z \over 2} C_{L,q}^{(S)(1)}\left({1 \over z}\right)
%\nonumber\\[2ex]
   +Z^{(T)(1)}_{gg}\otimes {C_F \over 2 N_f T_f} \left \{-{z \over 2}\,
C_{L,g}^{(1)S}\left({1 \over z}\right)\right \}~.  \nonumber
\end{eqnarray}
%------------------------------------------------------------------------
The transformations for the other NNLO coefficient 
functions~\cite{ZNR,ZN1} are given in Ref.~\cite{BRV1}.

Let us define the difference between the time--like physical
evolution kernel and the analytically continued space--like
evolution kernel by
%------------------------------------------------------------------------
\begin{eqnarray}
\delta K_{IJ} := K_{IJ}^T - \overline{K}_{IJ}^S.
\end{eqnarray}
%------------------------------------------------------------------------
Using identities for pair--convolutions of higher order Nielsen 
integrals, see Ref.~\cite{BRV1}, one finally obtains that
%------------------------------------------------------------------------
\begin{eqnarray}
\delta K_{d2} &=& 0 \nonumber\\
\delta K_{dd} &=& 0~,
\end{eqnarray}
%------------------------------------------------------------------------
and
%------------------------------------------------------------------------
\begin{eqnarray}
 \delta K_{22}^{N(1)} &=& 0 \nonumber\\
 \delta K_{L2}^{N(1)} &=& 0 \nonumber\\
 \delta K_{2L}^{N(1)} &=& 0 \nonumber\\
 \delta K_{LL}^{N(1)} &=& 0~,
\end{eqnarray}
%------------------------------------------------------------------------
showing explicitely the validity of the Drell--Levy--Yan relation for 
the physical evolution kernels in leading and next--to--leading order at
the above choice of observables.

We finally would like to comment on a relation suggested by Gribov and
Lipatov~\cite{GL},
%------------------------------------------------------------------------
\begin{eqnarray}
\overline{K}(x_E,Q^2) =  K(x_B,Q^2) \nonumber
\end{eqnarray}
%------------------------------------------------------------------------
This relation holds for the LO non-singlet contributions and some
pieces in the NLO non-singlet contributions, but is generally violated
{beyond LO.}
%===============================================================================
\section{Conclusions}
%===============================================================================

\vspace{1mm} \noindent
The scale evolution of {structure} and {fragmentation} functions
can be represented in terms of physical evolution kernels and
{observable}  non-perturbative input distributions.
The physical evolution kernels of either choice of observables
are related for the evolution of {structure} and {fragmentation}
functions by an analytic continuation {(DLY relation)} from 
{$0 \leq x < 1$} to {$1 < x < \infty$} up to {$O(\alpha_s^2)$},
for which transformation rules were derived.
The {Gribov--Lipatov relation} is violated beyond LO.
An extension of the present investigation to    {$O(\alpha_s^3)$}
requires the knowledge of the hitherto unknown    {3--loop} singlet
anomalous dimensions. The DLY relation for the evolution kernels is not
necessarily expected to hold to arbitrary high orders due to the
emergence of new production thresholds for the s-channel process.
An interesting test of QCD can be carried out in comparing the
scaling violations of {structure} and {fragmentation functions}
using factorization scheme--independent evolution equations.
%===============================================================================

\Acknowledgments
We would  like to thank the organizers of the conference, H. Haber and 
S.J. Brodsky, for having organized a stimulating meeting.
%%%%%%%%%%%%%%%%%%%%%%%%%%%%%%%%%%%%%%%%%%%%%%%%%%%%%%%%%%%%%%%%%%%%%%%%%%%%%

%%%%%%%%%%%%%%%%%%%%%%%%%%%%%%%%%%%%%%%%%%%%%%%%%%%%%%%%%%%%%%%%%%%%%%%%%%%%%%%
%%%%%%%%%%%%%%%%%%%%%%%%%%%%%%%%%%%%%%%%%%%%%%%%%%%%%%%%%%%%%%%%%%%%%%%%%%%%%%%
\end{document}